\documentclass[twocolumn]{jpsj3}
\def\v#1{\mib #1}
\def\EHal{{\tilde{E}}}
\def\epsgtilde{{\tilde{\epsilon}_{\rm G}}}

\def\runauthor{K. {\sc Hida}, K. {\sc Takano}, and  H. {\sc Suzuki}}
\title
{
Quantum Phase Transitions in Alternating-Bond Mixed Diamond Chains with Spins $1$ and $1/2$}
\author
{
Kazuo {\sc Hida}\thanks{E-mail address: hida@phy.saitama-u.ac.jp}, 
Ken'ichi {\sc Takano}$^{1}$ and  
Hidenori {\sc Suzuki}$^{1}$\thanks{Present address: Department of Physics, College of Humanities and Sciences, Nihon University, Setagaya-ku, Tokyo 156-8550, JAPAN}
}

\inst
{Division of Material Science, Graduate School of Science and Engineering, \\ Saitama University, Saitama, Saitama, 338-8570, JAPAN\\ 
$^{1}$Toyota Technological Institute, 
Tenpaku-ku, Nagoya 468-851, JAPAN}

\recdate
{
November 20,  2009}

\abst
{
We investigate the mixed diamond chain composed of spins $1$ and $1/2$ when 
the exchange interaction is alternatingly distorted. 
Depending on the strengths of frustration and distortion, 
this system has 
 various ground states. Each ground state consists of 
an array of spin clusters separated by singlet dimers by virtue of an infinite number of local conservation laws.
We determine the ground-state phase diagram by numerically analyzing each spin cluster. 
In particular, for strong distortions, we find  
an infinite series of quantum phase transitions 
using the cluster expansion method and conformal field theory. 
This leads to an infinite series of steps in the  
 behavior of  Curie constant and residual entropy. 
}

\kword
{mixed diamond chain, bond alternation, 
spin cluster, 
infinite series of transitions, quantum phase transition 
}

\begin{document}
%\sloppy
\maketitle
\section{Introduction}

Frustration plays a crucial role in low-dimensional quantum magnetism.\cite{hfm2008,diep} It not only drives  magnetically ordered states into disordered states by enhancing quantum fluctuation but also induces magnetic moments from the disordered phase. Remarkably, there exist a class of models whose ground states are exactly written down as  spin cluster states (SCSs) because of frustration.  
A SCS is a tensor product of exact local eigenstates of cluster spins; 
a dimer state is a special case.

 One of the well-known examples is the Majumdar-Ghosh model, which is a spin 1/2 antiferromagnetic Heisenberg chain with next-nearest-neighbor interaction whose magnitude is half of the nearest-neighbor interaction.\cite{mg} This ground state is a prototype of spontaneously dimerized phases in one-dimensional frustrated %one dimensional 
magnets. Many corresponding materials are also found as listed in ref. \citen{hase}. 
Another well-known example is the Shastry-Sutherland model\cite{ss} for which a corresponding material was synthesized 18 years after its theoretical prediction.\cite{kage1,kage2}  The ground states of these models are, however, nonmagnetic. Namely, the possible magnetic (quasi-) long-range order in the unfrustrated counterpart of these models is destroyed by the enhancement of quantum fluctuation by frustration.

One of the  authors and coworkers investigated 
a diamond chain consisting of the same kind of spin, i.e., a pure diamond chain (PDC), as an exactly treatable frustrated model.\cite{Takano-K-S} 
They found that, in the spin-1/2 case, 
it has two different SCSs as the ground state depending on the strength of frustration. 
One of them is the nonmagnetic phase accompanied by the spontaneous translational symmetry breakdown (STSB) and the other is a paramagnetic phase without STSB. 
It is also found that this model has 
a ferrimagnetic ground state 
in a less frustrated region.

Modifications of the PDC have been examined 
by many authors. Among them, the PDC with distortion 
has been thoroughly investigated by numerical methods.\cite{ottk,otk,sano} 
It is found that a natural mineral azurite consists of distorted PDCs with spin 1/2. The magnetic properties of this material have been experimentally studied in detail.\cite{kiku,ohta} 
Other materials have also been reported.\cite{izuoka,uedia} 
The effects of the 
4-spin cyclic interaction have recently been  investigated  by Ivanov {\it et al.}\cite{dia4spin} It should be remarked that the diamond chain structure is one of the simplest structures compatible with the 4-spin cyclic interaction. 
The finite temperature properties of diamond chains consisting of Heisenberg bonds and Ising bonds are investigated exactly.\cite{str1,str2} The thermodynamic properties of a similar classical model with a hierarchical structure are also studied.\cite{fuku} The behavior of the residual entropy of this model is similar to that of the diamond chain.

\begin{figure}[t] % Fig. 1
%\centerline{\includegraphics[width=6cm]{lattice_ladder_plaq.eps}}
\centerline{\includegraphics[width=6cm]{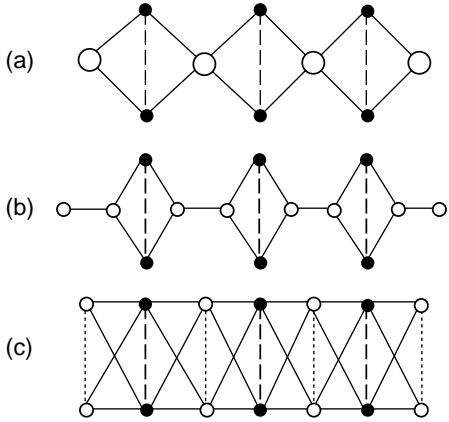}}
\caption{Structure of (a) the mixed diamond chain and related models: (b) the dimer plaquette chain, (c) the frustrated ladder. The large open circles are spin-1 sites, The small open circles and filled circles are spin-1/2 sites.}
\label{lattice_lp}
\end{figure}

 In our previous work\cite{tsh}, we introduced another version of the diamond chain that consists of two kinds of spins, and named it the mixed diamond chain (MDC). 
Although the MDC is a simple extension of the PDC, 
it is a model belonging 
to a 
class different from that of the PDC in that 
its ground state is the Haldane state in the weak frustration region. We have investigated  the MDC consisting of spins 1 and 1/2 depicted in Fig. \ref{lattice_lp}(a) in special detail.\cite{tsh,hts}  
Since the MDC has an infinite number of local conservation laws like the PDC, a typical ground state  is a SCS 
  consisting of spin clusters each carrying 
  spin 1.  
A series of quantum phase transitions take place  between 
 five ground-state phases with different periodicities and with or 
   without a STSB.  
Except in the Haldane phase, the MDC has macroscopically degenerate ground states, 
because each cluster has three degenerate ground states with spin 1. The SCS structures of the ground states 
are reflected in the characteristic thermal properties, as reported in ref. \citen{hts}.

 %=====================================
%=====================================
The MDC is related to other important models of frustrated quantum magnetism. 
The dimer-plaquette model\cite{plaq,plaq2} shown in Fig. \ref{lattice_lp}(b), which may be regarded as a one-dimensional counterpart of the Shastry-Sutherland model, reduces to the MDC if the horizontal bond in Fig. \ref{lattice_lp}(b) is strongly ferromagnetic. The ground state of the dimer-plaquette model in this region has not yet been studied.  In the MDC limit, however, it becomes clear that the ground state is the Haldane state. In addition, it turns out that {\it all} eigenstates can be expressed in terms of the eigenstates of  finite-length spin 1 Heisenberg chains for an arbitrary strength of the plaquette bond in the MDC limit.\cite{tsh} 
 
 The MDC is also related to rung-alternating frustrated Heisenberg ladders (Fig. \ref{lattice_lp}(c)). If the spin 1 site in MDC is decomposed into the sum of two spin 1/2's, the MDC is equivalent to the low-energy sector of the spin 1/2 rung alternating frustrated ladder with a leg coupling equal to the diagonal coupling and one of the rung couplings is weakly antiferromagnetic or ferromagnetic. In the absence of rung alternation, it is known that this model has only Haldane and rung-dimer ground states.\cite{frulad1} However, in the MDC limit, the intermediate phases appear as explained above.  

Thus far, no materials described by the MDC model have been found. Nevertheless, synthesizing MDC materials is not an unrealistic expectation in view of the success of the synthesis of many low-dimensional bimetallic magnetic compounds\cite{m-d} 
 and organic magnetic compounds. In the latter, spin 1 units 
 included in a MDC material can  be formed as ferromagnetic dimers.\cite{cb} 
Generally, materials with a lower symmetry have a higher chance of being found or synthesized than those with a higher symmetry. 
Therefore, to raise the possibility of realizing 
 MDC materials, theoretical predictions on modified versions of the MDC model with a low symmetry are required.

%=====================================
\begin{figure}[t] % Fig. 2
%\centerline{\includegraphics[width=6cm]{mode.eps}}
\centerline{\includegraphics[width=6cm]{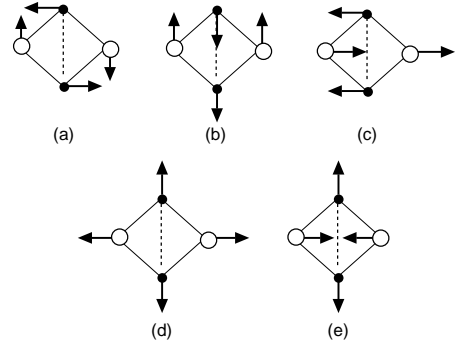}}
\caption{Displacement modes of a diamond unit.}
\label{mode}
\end{figure}
%=====================================
The distortion is a realistic modification that lowers the symmetry of the MDC. 
To classify distortion patterns, we consider the normal modes of a diamond  unit. It has 8 degrees of freedom of displacement within the diamond plane. Excluding two translations and one rigid body rotation, we have five normal modes as depicted in Fig. \ref{mode}. We may also expect that the distorted MDC can be  realized as a result of the collective softening of this normal mode. Two of them ((a) and (b)) break the local conservation laws, which hold in undistorted MDCs. 
Among other distortion patterns ((c), (d) and (e)), 
(d) and (e) do not change the geometry of the original undistorted MDC. 
The distortion pattern (c) brings about the bond alternation in the undistorted MDC to break the original reflection symmetry. 
This type of distortion not only raises the possibility of the experimental realization of the MDC, but also is of theoretical importance, because in this case 
the eigenstates are exactly expressed as SCSs by virtue of
  the same local conservation laws as in the undistorted case. 
We hence restrict ourselves to this case in the present paper. 
We will show that the bond alternation distortion does not destroy the exotic phases of the undistorted MDC but even makes the ground-state phase diagram 
 more exotic than that of the undistorted MDC. 
Since cases (a) and (b) contain different physics owing to 
 effective interactions between spin clusters, they will be reported in a separate paper.%\cite{htstag}

This paper is organized as follows. 
 {In \S 2}, the model Hamiltonian is presented. 
 {In \S 3}, the classical limit is discussed. 
 {In \S 4}, the structure of the ground state is explained and the ground-state phases are numerically determined.  
In \S 5, an infinite series of phase transitions that occur for strong 
bond alternation are examined using numerical, cluster expansion, and conformal field theory methods. 
The last  section is devoted to summary and discussion.

\section{Hamiltonian}
%\label{section:ham}

\begin{figure}[t] % Fig. 1 --> 3
%\centerline{\includegraphics[width=6cm]{lattice_alt.eps}}
\centerline{\includegraphics[width=6cm]{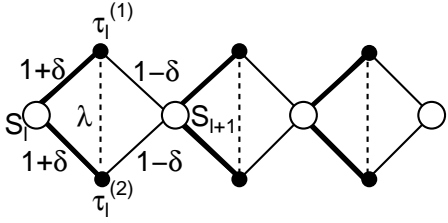}}
\caption{Structure of alternating bond mixed diamond chain with $S=1$ and $\tau^{(1)}=\tau^{(2)}=1/2$.
 $\lambda$, $1+\delta$, and $1-\delta$ are the exchange parameters, 
where $\lambda$ and $\delta$ control 
 the strengths 
  of frustration and alternating distortion, respectively.}
\label{lattice}
\end{figure}

 The alternating bond MDC  is described by the Hamiltonian 
%------------------------------------------------------------
\begin{align}
{\mathcal H} = &\sum_{l=1}^{N} \left\{(1+\delta)\v{S}_{l}(\v{\tau}^{(1)}_{l}+\v{\tau}^{(2)}_{l}) \right. 
\nonumber\\
&+(1-\delta)(\v{\tau}^{(1)}_{l}+\v{\tau}^{(2)}_{l})\v{S}_{l+1}
+ \left. \lambda\v{\tau}^{(1)}_{l}\v{\tau}^{(2)}_{l}\right\} , 
\label{hama}
\end{align}
%------------------------------------------------------------
where $\v{S}_{l}$ and $\v{\tau}^{(\alpha)}_{l}\ (\alpha=1,2)$ are spin 1 and spin 1/2 operators, respectively.  The number of the unit cells is denoted by $N$. 
 The Hamiltonian includes three types of exchange parameters, namely, 
$1+\delta$, $1-\delta$, and $\lambda$;  
$\delta$ represents the strength of the alternating-bond distortion, and $\lambda$  controls  the frustration as depicted in Fig. \ref{lattice}. 
In the case of $\delta = 0$, eq.~(\ref{hama})  reduces to 
the Hamiltonian  of 
the MDC without distortion.\cite{tsh} 

%=====================================
%=====================================

\section{Classical Ground States}

Before analyzing the full quantum system (\ref{ham2}), 
we examine its classical version 
 in which the spin operators $\v{S}_{l}$ and $\v{\tau}_{l}$ are replaced by classical vectors with the length $S$ and $\tau$, respectively.  We also impose the condition 
 $\tau = S/2$. 
By introducing ${\tilde{\v{S}}}_{l} \equiv (1+\delta){\v{S}}_{l} + (1-\delta){\v{S}}_{l+1}$ and 
$\v{T}_{l} \equiv \v{\tau}^{(1)}_{l}+\v{\tau}^{(2)}_{l}$, 
 the classical  Hamiltonian 
 is expressed in the following two forms: 
%------------------------------------------------------------
\begin{align} %
{\mathcal H}^{\mathrm{cl}} 
&= \frac{1}{4} \sum_l 
\left[ (2{\v{T}}_{l}+{\bf{\tilde S}}_{l})^{2} 
- 2(2 -\lambda) {\v{T}}_{l}^{2} 
- {\bf{\tilde S}}_{l}^{2} - \lambda S^2 \right], 
\label{classic1}\\
&= \frac{1}{4\lambda} \sum_l 
\left[2(\lambda{\v{T}}_{l}+{\tilde{\v{S}}}_{l})^{2} 
- 2{\tilde{\v{S}}}_{l}^{2} - \lambda^2 S^2\right]. 
\label{classic2}
\end{align}
%------------------------------------------------------------
 These equations  have the same form  
as those in the absence of distortion,\cite{tsh} since the distortion parameter is absorbed in ${\tilde{\v{S}}}_{l}$. 

For $\lambda \le 2$, the expression (\ref{classic1}) shows that 
${\mathcal H}^{\mathrm{cl}}$ is minimized  
if $|{\tilde{\v{S}}}_{l}|=2S$, $|{\v{T}}_{l}|=S$ 
and $|{\v{T}}_{l}+\frac{1}{2}{\tilde{\v{S}}}_{l}|=0$. 
Then all the ${\v{S}}_{l}$'s ($\v{\tau}_{l}^{(\alpha)}$'s) 
in the chain are aligned parallel (antiparallel) to a fixed axis, 
and the ground state is antiferromagnetic. 
This ground state is elastic, since any local modification of 
the spin configuration increases the energy. 
For $\lambda > 2$, the expression (\ref{classic2}) reveals that 
${\mathcal H}^{\mathrm{cl}}$ is minimized  
if $|{\tilde{\v{S}}}_{l}|=2S$ 
and $|{\v{T}}_{l}+{\tilde{\v{S}}}_{l}/\lambda |=0$. 
Hence, ${\v{S}}_{l}$ and ${\v{S}}_{i+1}$ are parallel, and 
$\v{\tau}_{l}^{(1)}$ and $\v{\tau}_{l}^{(2)}$ 
form a triangle with ${\tilde{\v{S}}}_{l}/\lambda$. 
$\v{\tau}_{l}^{(1)}$ and $\v{\tau}_{l}^{(2)}$ 
may be rotated about 
the axis of ${\v{S}}_{l}$ and ${\v{S}}_{i+1}$ 
without raising the energy. 
Then all the ${\v{S}}_{l}$'s in the chain are aligned 
parallel to a fixed axis, and the arbitrariness of the local rotation 
of $\v{\tau}_{l}^{(1)}$ and $\v{\tau}_{l}^{(2)}$ is not obstructed. 
Thus, the ground state is ferrimagnetic 
with magnetization $(1 - 2/\lambda)SN$. 

Thus, we have two classical phases separated by the phase boundary $\lambda = 2$ independent of $\delta$. 
In fact, $\delta$ is embedded in ${\tilde{\v{S}}}_{l}$ and 
does not explicitly appear in the energy expressions eqs. (\ref{classic1}) and (\ref{classic2}). 
The distortion affects the ground state only in the presence of quantum effect. 
As was pointed out, for $\lambda > 2$, the classical ground-state configuration can be locally modified with no energy increase. 
This classical situation corresponds to a quantum situation in which 
there are an infinite number of low-energy states that are made from each other by local modification. 
Since such quasi-degenerate low-energy states may enhance quantum fluctuations, we expect the appearance of exotic quantum states  {for $\lambda \gtrsim 2$} in the quantum system (\ref{ham2}).

\section{Ground-State Phase Diagram}

The Hamiltonian (\ref{hama}) has a series of conservation laws. 
To see it, we rewrite eq. (\ref{hama}) 
 in the form, 
%------------------------------------------------------------
\begin{align}
{\mathcal H} = \sum_{l=1}^{N} \left[(1+\delta)\v{S}_{l}\v{T}_{l}+(1-\delta)\v{T}_{l}\v{S}_{l+1}
+ \frac{\lambda}{2}\left(\v{T}^2_{l}-\frac{3}{2}\right)\right] . 
\label{ham2}
\end{align}
%------------------------------------------------------------
where the composite spin operators $\v{T}_l$ are defined as 
%------------------------------------------------------------
\begin{align}
\v{T}_{l} \equiv \v{\tau}^{(1)}_{l}+\v{\tau}^{(2)}_{l} 
\quad (l = 1, 2, \cdots N). 
\end{align}
%------------------------------------------------------------
Then it is  evident that 
%------------------------------------------------------------
\begin{align}
[\v{T}_l^2, {\mathcal H}] = 0 \quad (l = 1, 2, \cdots N). 
\end{align}
%------------------------------------------------------------
Thus, we have $N$ conserved quantities $\v{T}_l^2$ for all $l$, even if we have introduced an alternating-bond distortion. 
By defining the magnitude $T_l$ of the composite spin $\v{T}_l$ by $\v{T}_l^2 = T_l (T_l + 1)$, we have  a 
 set of good quantum numbers $\{T_l; l=1,2,...,N\}$. 
Each $T_l$ takes a value of 0 or 1. 
The total Hilbert space of eq.~(\ref{ham2}) consists of 
separated subspaces, each of which is specified by 
a definite set of $\{T_l\}$, i.e., a sequence of 0 and 1. 
 A pair of spins with $T_l=0$ is a singlet dimer. 
 A cluster including $n$ successive $T_l=1$ pairs bounded by two $T_l=0$ pairs is called a cluster-$n$. 

 Thus,  the eigenstates of a cluster-$n$  are expressed  in terms of   
the spin states of $n$ $\v{T}_{l}$'s with $T_l=1$ and $n+1$ $\v{S}_{l}$'s of $S_l=1$. 
Hence, a cluster-$n$ is equivalent to 
an alternating bond antiferromagnetic Heisenberg chain (AAFH1)  consisting of $2n+1$ effective spins with spin magnitude 1  as in
 the undistorted case.\cite{tsh,hts} 
We follow the terminology of refs. \citen{tsh} and \citen{hts} to call a ground state consisting of a uniform array of cluster-$n$'s as  the dimer-cluster-$n$ (DC$n$) phase. 
The phase boundary between  the DC$(n-1)$ and  DC$n$ phases is given by
%------------------------------------------------------------
\begin{align}
\lambda_{\rm c}(n-1,n; \delta) =& (n+1)\EHal_{\rm G}(2n-1,\delta)
-n\EHal_{\rm G}(2n+1,\delta), 
\label{bdry}
\end{align}
%------------------------------------------------------------
where $\EHal_{\rm G}(L,\delta)$ is the ground-state energy of  the AAFH1 with a bond alternation  
 $\delta$ and a length $L$. These phase transitions are of the first order, since they take place as level crossings between two eigenstates of the Hamiltonian (\ref{hama})
 characterized by different sets of quantum numbers $\{T_l\}$. 
 
The width $\Delta\lambda(n {; \delta})$ of the DC$n$ phase  for each $\delta$ is given by 
%------------------------------------------------------------
\begin{align}
\Delta\lambda(n {; \delta})&=\lambda_{\rm c}(n-1,n; \delta)-\lambda_{\rm c}(n,n+1; \delta)\\
&=(n+1)\Bigl[\EHal_{\rm G}(2n-1,\delta)-2\EHal_{\rm G}(2n+1,\delta)\Bigr.\nonumber\\
&\Bigl. \qquad\qquad\qquad +\EHal_{\rm G}(2n+3,\delta)\Bigr] . 
\label{width}
\end{align}
%------------------------------------------------------------
 Equation (\ref{width}) shows that $\Delta\lambda(n {; \delta})$  is proportional to the second difference of  $\EHal_{\rm G}(2n+1,\delta)$ with respect to $n$. 
Therefore, the DC$n$ phase appears with a finite width as long as $\EHal_{\rm G}(2n+1,\delta)$ is a convex function of $n$. If $\Delta\lambda(n {; \delta})$ is positive for all $n$, an infinite series of phase transitions take place before reaching the DC$\infty$ phase. 

%=====================================
\begin{figure} % Fig. 2(a)(b)(c) --> 4
%\centerline{\includegraphics[width=7cm]{eneg_ndep_hal.eps}}
%\centerline{\includegraphics[width=7cm]{eneg_ndep_dim.eps}}
%\centerline{\includegraphics[width=7cm]{eneg_ndep_dmrg.eps}}
\centerline{\includegraphics[width=7cm]{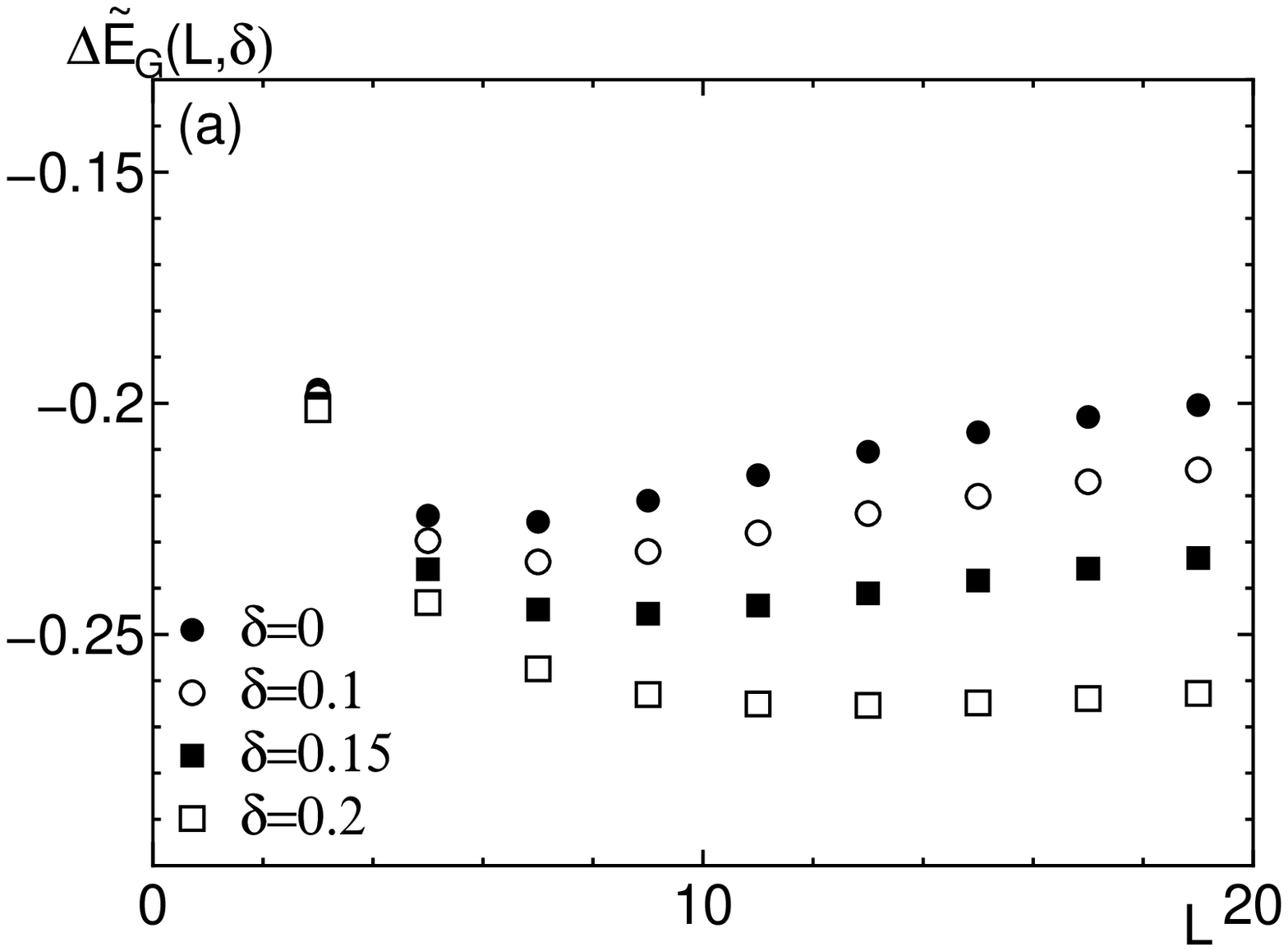}}
\centerline{\includegraphics[width=7cm]{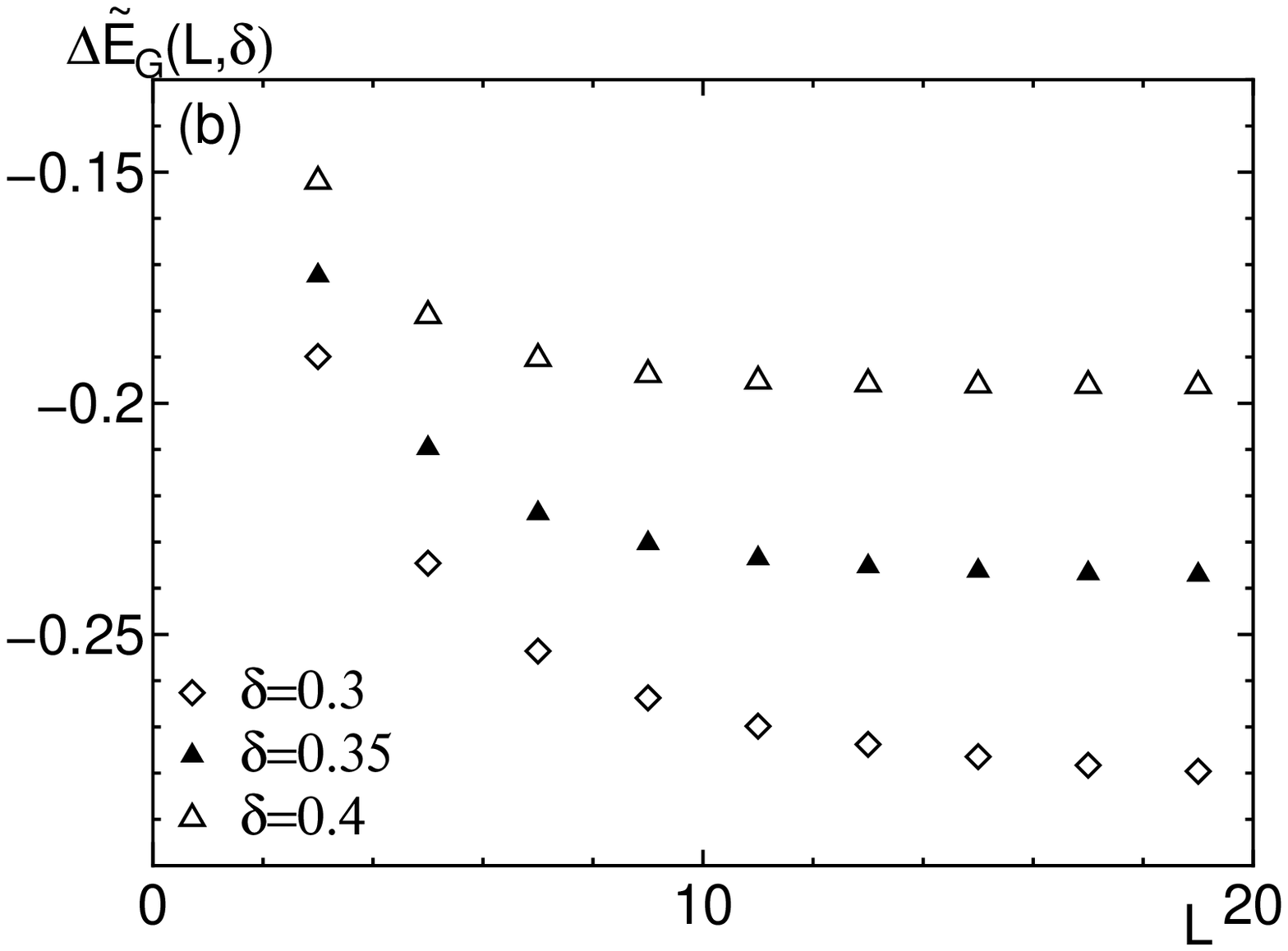}}
\centerline{\includegraphics[width=7cm]{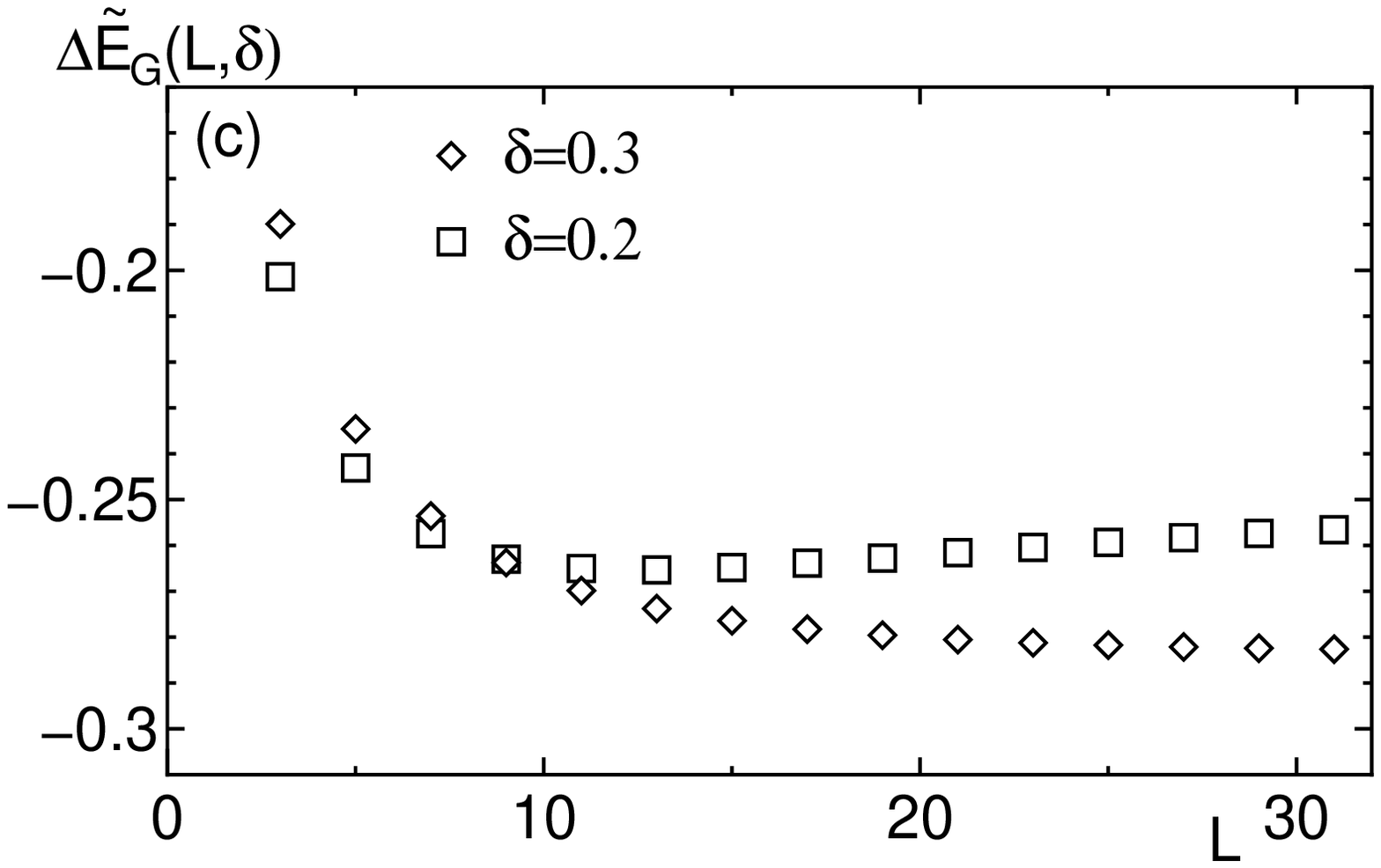}}
\caption{Size dependence of the ground-state boundary energy $\Delta\EHal_{\rm G}(L,\delta)$ of AAFH1 for (a) $0 < \delta < \delta_{\rm c}$ and  (b) $\delta > \delta_{\rm c}$ obtained by numerical diagonalization, and that for (c) $\delta=0.2$ and 0.3 obtained by DMRG.}
\label{eg}
\end{figure}
%=====================================

We have calculated $\EHal_{\rm G}(L,\delta)$ by   numerical diagonalization for $3 \leq L \leq 19 $ and  by  the finite-size density matrix renormalization group (DMRG) for larger $L$. 
 To extract the boundary contribution, we subtract the  corresponding bulk  energy 
 from $\EHal_{\rm G}(L,\delta)$ as 
%------------------------------------------------------------
\begin{align}
\Delta\EHal_{\rm G}(L,\delta)\equiv\EHal_{\rm G}(L,\delta)-(L-1)\epsgtilde(\infty,\delta), 
\end{align}
%------------------------------------------------------------
where $\epsgtilde(\infty,\delta)$ is the ground-state energy of the infinite-size AAFH1 per spin.  This subtraction does not influence the second difference. 
The system size dependence of $\Delta\EHal_{\rm G}(L,\delta)$ is shown in Figs. \ref{eg}(a) and \ref{eg}(b) for $3 \leq L \leq 19$. 
Figure \ref{eg}(c) shows the data for $L \leq 31$ for $\delta=0.3$ and 0.2. 
The energy $\epsgtilde(\infty,\delta)$ is estimated using the infinite-size DMRG by measuring the bond energy of the two bonds closest to the center of the chain. 

Figure \ref{eg} shows that the ground-state energy of  the AAFH1 is a 
 convex function of $L$ for large $\delta$ within numerical data. 
Therefore, DC$n$ phases are realized for all $n$ depending on $\lambda$. 
In contrast, for small $\delta$, 
 no DC$n$ phases with large $n$  appear, 
since the ground-state energy of  the AAFH1 is a concave function of $L$ for large $L$ as shown in Fig. \ref{eg}. 
Actually, it is known that no DC$n$ phases with $n \geq 4$  appear for $\delta=0$.\cite{tsh} 
In such cases, the direct transition from the DC$n$ phase to the DC {$\infty$} phase takes place at 
%------------------------------------------------------------
\begin{align}
\lambda_{\rm c}(n,\infty {; \delta})&=\EHal_{\rm G}(2n+1,\delta)- {2}(n+1)\epsgtilde(\infty,\delta), % These values are checked on 2008.9.15
\label{bdryinf}
\end{align}
%------------------------------------------------------------
if $\lambda_{\rm c}(n,\infty {; \delta}) < \lambda_{\rm c}(n,n+1 {; \delta})$.

\begin{figure} % Fig. 3 --> 5
%\centerline{\includegraphics[width=7cm]{craltphase.eps}}
\centerline{\includegraphics[width=7cm]{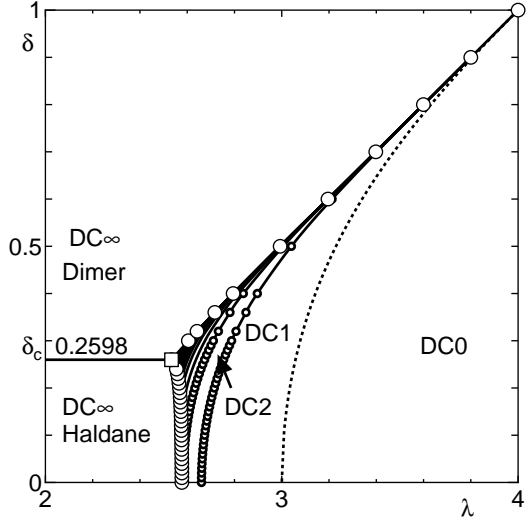}}
\caption{Ground-state phase diagram. Small open circles are the critical points $\lambda_{\rm c}(n,n-1;\delta)$ and large open circles are the critical points  $\lambda_{\rm c}(\infty,n;\delta)$. The solid lines are guides for the eye. For $n \geq 4$, phase boundaries are only shown by solid lines to avoid complication. The values of $\lambda_{\rm c}(0,1)$ calculated using eq. (\ref{analamc}) are shown by the thick dotted line. The open square is  $\lim_{n\rightarrow\infty}\lambda_{\rm c}(n,n-1;\delta_{\rm c})$. }
\label{phase}
\end{figure}

 The phase diagram is shown in Fig. \ref{phase} using these values. At $\delta=\delta_{\rm c}\simeq 0.2598$, where the Haldane-dimer phase transition takes place in the infinite-size AAFH1\cite{kn}, the convergence of the infinite-size DMRG becomes worse.  
 Hence, we employed  extrapolation from the finite-size DMRG results to determine  $\epsgtilde(\infty,\delta)$. For $\delta < \delta_{\rm c}$, the DC$\infty$ ground state corresponds to the uniform Haldane phase, while it corresponds to the dimer phase for $\delta > \delta_{\rm c}$.

 The phase boundary between the DC0 and DC1 phases is analytically obtained. 
Obviously, $\EHal_{\rm G}(1,\delta)=0$, and $\EHal_{\rm G}(3,\delta)$ satisfies the following eigenvalue equation within the subspace with total spin 1:
%------------------------------------------------------------
\begin{align}
\EHal_{\rm G}(3,\delta)^3+4\EHal_{\rm G}(3,\delta)^2+(3-7\delta^2)\EHal_{\rm G}(3,\delta)-16\delta^2=0 . 
\end{align}
%------------------------------------------------------------
Therefore, $\lambda_{\rm c}(0,1)$ satisfies
%------------------------------------------------------------
\begin{align}
\lambda_{\rm c}(0,1 {; \delta})^3 
&-4\lambda_{\rm c}(0,1 {; \delta})^2+(3-7\delta^2)\lambda_{\rm c}(0,1 {; \delta}) \nonumber\\
&+16\delta^2=0.
\label{analamc}
\end{align}
%------------------------------------------------------------
This relation is 
plotted in Fig. \ref{phase} by the thick dotted line. For $\delta \simeq 1$, eq. (\ref{analamc}) implies $\lambda_{\rm c}(0,1 {; \delta}) \simeq 4-2(1-\delta)$. 

%=====================================
%=====================================

\section{Infinite Series of Quantum Phase Transitions}  

\subsection{Almost decoupled limit: $\delta\simeq 1$}
\begin{figure} % Fig. 4 --> 6
%\centerline{\includegraphics[width=7cm]{widthratio.eps}}
\centerline{\includegraphics[width=7cm]{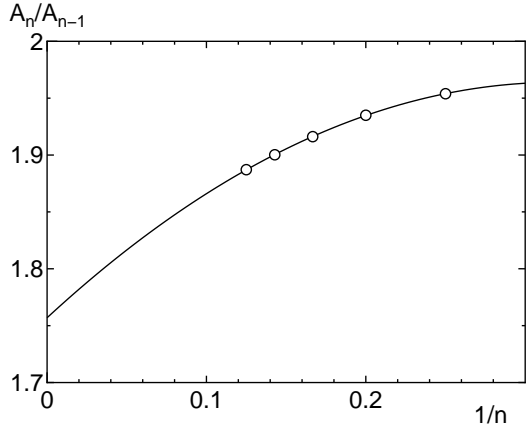}}
\caption{Ratio $r_n\equiv A_n/A_{n-1}$ plotted against $1/n$. The solid line is the least-squares fit by the second-order polynominal of $1/n$.}
\label{ratio}
\end{figure}

At $\delta=1$, an  AAFH1 with a length of $2n+1$ is decoupled into $n$ independent dimers and a single spin 1. Therefore, we can employ the cluster expansion method\cite{gsh} to estimate $\EHal_{\rm G}(L,\delta)$. 
The lowest-order nonvanishing  contribution  
to $\Delta\lambda(n {; \delta})$  is 
$O((1-\delta)^{2n})$  and is written as 
%------------------------------------------------------------
\begin{align}
\Delta\lambda(n {; \delta})&\simeq A_n(1+\delta)\left(\frac{1-\delta}{1+\delta}\right)^{2n}
\simeq \frac{A_n}{2^{2n-1}}\left({1-\delta}\right)^{2n} . 
\label{eq:dimerwidth}
\end{align}
%------------------------------------------------------------
The factor  {$A_n$ is given by} 
%------------------------------------------------------------
\begin{align}
    A_1&=    4/3 , \nonumber\\
    A_2&=    2.305555556 , \nonumber\\
    A_3&=    4.501164348 , \nonumber\\
    A_4&=    8.794932620 , \nonumber\\
    A_5&=    17.01798127 , \nonumber\\
    A_6&=    32.60950955 , \nonumber\\
    A_7&=    61.96667474 , \nonumber\\
    A_8&=    116.9402383,
\end{align}
%------------------------------------------------------------    
up to $n=8$. The ratio $r_n\equiv A_n/A_{n-1}$ is plotted against $1/n$ in Fig. \ref{ratio}. This ratio tends to converge to $r_{\infty}\simeq 1.76$ and does not oscillate with $n$. 
 Hence, we speculate that
$A_n > 0$ for all $n$. This implies that an infinite series of phase transitions take place for $\delta \simeq 1$. 
%=====================================
%=====================================

\subsection{$\delta=\delta_{\rm c}$}
Because the infinite series of phase transitions are most prominent around $\delta=\delta_{\rm c}$ in Fig. \ref{phase}, we investigate this point in more detail. 
At this point, the ground state of  the infinite AAFH1 is on the Gaussian critical point described by the conformal field theory with  a conformal charge  $c=$1. 
The ground state of  a cluster-$n$ is also described as a finite-size AAFH1 at the Gaussian critical point. It is known that the finite-size ground-state energy of the Gaussian model with an open boundary condition behaves as\cite{bcn} 
%------------------------------------------------------------
\begin{align}
\EHal_{\rm G}(L,\delta_{\rm c})=L\epsgtilde(\infty,\delta_{\rm c})+\epsilon_{\rm b}+\frac{A}{L}, \ A=\frac{\pi v c}{24},
\label{eq:asym}
\end{align}
%------------------------------------------------------------
for large $L$, where $L$ is the length of the chain, $\epsilon_{\rm b}$ is the boundary energy, $v$ is the spin wave velocity, and $c$ $(=1)$ is the conformal charge.

Fitting the ground-state energy obtained using the finite-size DMRG for $21 \leq L \leq 55$, we obtain
%------------------------------------------------------------
\begin{align}
\epsgtilde(\infty,\delta_{\rm c})&= -1.429085\pm 0.000005 , \nonumber \\
 \epsilon_{\rm b}&= 1.1045\pm 0.0005 , 
 \nonumber\\
   A&=0.41 \pm 0.005 , 
\end{align}
%------------------------------------------------------------
where errors are estimated by changing the smallest system size used for the fitting from 21 to 29. 

Substituting  eq. (\ref{eq:asym}) in eq. (\ref{bdry}), we obtain
%------------------------------------------------------------
\begin{align}
\lambda_{\rm c}(n-1,n; \delta_{\rm c})&=-\epsgtilde(\infty,\delta_{\rm c})+\epsilon_{\rm b}+A\left(\frac{4n+1}{4n^2-1}\right)\nonumber\\
&\simeq  2.5335+ 0.41\left(\frac{4n+1}{4n^2-1}\right) . 
\label{eq:asymlc}
\end{align}
%------------------------------------------------------------
The width of the DC$n$ phase is given by
%------------------------------------------------------------
\begin{align}
\Delta\lambda(n {; \delta})&=A\left(\frac{8(n+1)}{(2n-1)(2n+1)(2n+3)}\right)\nonumber\\
&\simeq 0.41\left(\frac{8(n+1)}{(2n-1)(2n+1)(2n+3)}\right).     
%\label{eq:asymwidth}
\end{align}
%------------------------------------------------------------
This expression is  positive for all $n \geq 1$.  Therefore, an infinite series of phase transitions take place at $\lambda_{\rm c}(n-1,n; \delta_{\rm c})$. It should be remarked that the width of the DC$n$ phase decreases with $n$ algebraically, while it decreases exponentially for $\delta \simeq 1$ according to eq. (\ref{eq:dimerwidth}). This explains why the infinite series of transitions are most visible at $\delta=\delta_{\rm c}$ in Fig. \ref{phase}.

\begin{figure} % Fig. 5 --> 7
%\centerline{\includegraphics[width=7cm]{lamdac_cr0.eps}}
\centerline{\includegraphics[width=7cm]{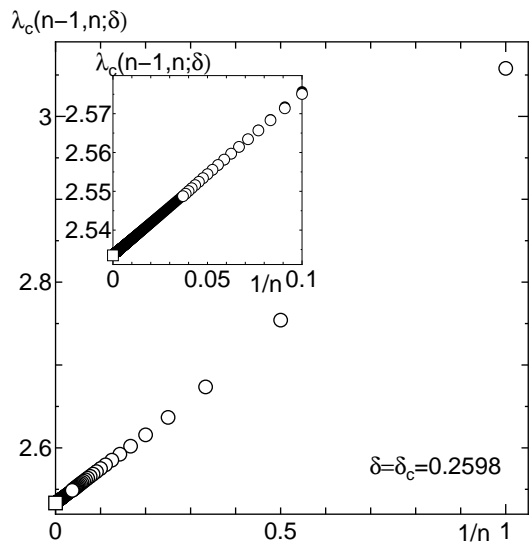}}
\caption{Infinite series of critical points $\lambda_{\rm c}(n-1,n;\delta_{\rm c})$ at $\delta=\delta_{\rm c}$. 
Open circles are calculated with eq. (\ref{bdry}) using the values of $\EHal_{\rm G}(2n+1,\delta_{\rm c})$ obtained by numerical diagonalization  and finit-size DMRG. 
Filled circles are obtained using eq. (\ref{eq:asymlc}). 
The open square is  $\lim_{n\rightarrow\infty}\lambda_{\rm c}(n,n-1;\delta_{\rm c})$. 
The inset is an enlarged figure for large $n$.}
\label{lambdac}
\end{figure}

We can also estimate $\lambda_{\rm c}(n-1,\infty; \delta_{\rm c})$ as
%------------------------------------------------------------
\begin{align}
\lambda_{\rm c}(n-1,\infty; \delta_{\rm c})&=-\epsgtilde(\infty,\delta_{\rm c})+\epsilon_{\rm b}+\frac{A}{2n-1},     
\end{align}
%------------------------------------------------------------
by substituting eq. (\ref{eq:asym}) in eq. (\ref{bdryinf}). Thus, we obtain
%------------------------------------------------------------
\begin{align}
\lambda_{\rm c}(n-1,n; \delta_{\rm c})-\lambda_{\rm c}(n-1,\infty; \delta_{\rm c})&=\frac{2nA}{4n^2-1}. 
\end{align}
%------------------------------------------------------------
This implies that $\lambda_{\rm c}(n-1,n; \delta_{\rm c})>\lambda_{\rm c}(n-1,\infty; \delta_{\rm c})$ for all $n \geq 1$. This confirms our numerical conclusion that no direct transition from the DC$n$ phase to the DC$\infty$ phase takes place. The critical values $\lambda_{\rm c}(n-1,n;\delta_{\rm c})$ are plotted against $n$  in Fig. \ref{lambdac}. 

%=====================================
%=====================================

% {
\subsection{Physical quantities}
Physical quantities in the ground state  is 
estimated in the same way as in the case of 
 the undistorted MDC.\cite{tsh,hts} Because each cluster-$n$ carries a spin 1 degree of freedom, the  low temperature magnetic susceptibility obeys the Curie law with the Curie constant $C$ per unit cell given by 
%------------------------------------------------------------
\begin{align}
C\simeq\frac{2}{3(n+1)},
\label{eq:curie}
\end{align}
%------------------------------------------------------------
and the residual entropy per unit cell is given by 
%------------------------------------------------------------
\begin{align}
S&=\frac{1}{n+1}\ln \, 3,
\label{eq:ent0}
\end{align}
%------------------------------------------------------------
in the DC$n$ phase.%}. 

%=====================================
\begin{figure} % Fig. 6 --> 8
%\centerline{\includegraphics[width=7cm]{curie_cr.eps}}
\centerline{\includegraphics[width=7cm]{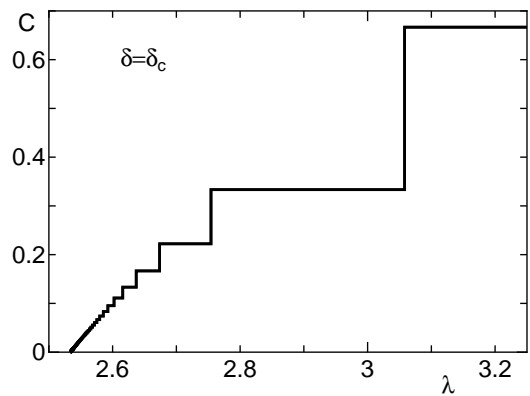}}
\caption{$\lambda$-dependence of Curie constant $C$ at $\delta=\delta_{\rm c}$.}
\label{curiec}
\end{figure}
%=====================================

%=====================================
\begin{figure} % Fig. 7 --> 9
%\centerline{\includegraphics[width=7cm]{entcr.eps}}
\centerline{\includegraphics[width=7cm]{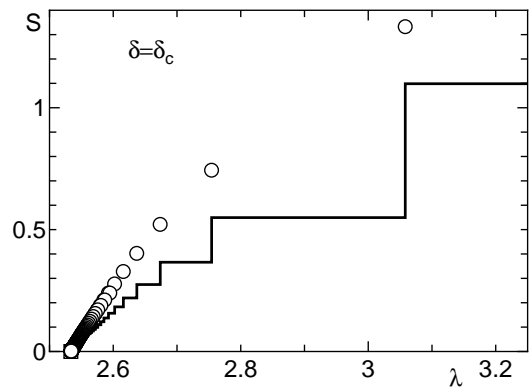}}
\caption{$\lambda$-dependence of residual entropy $S$ at $\delta=\delta_{\rm c}$. Open circles are values on the phase boundary.}
\label{entcr}
\end{figure}
%=====================================

On the phase boundary between the DC$(n-1)$ and DC$n$ phases,  the residual entropy is larger than those in the DC$n$ and DC$(n-1)$ phases owing to the mixing entropy of cluster-$n$'s and cluster-$(n-1)$'s. It is given by
%------------------------------------------------------------
\begin{align}
S = - \, \ln \frac{1-x}{x} , \label{eq:enta}
\end{align}
%------------------------------------------------------------
where $x$ is the solution of
%------------------------------------------------------------
\begin{align}
(n+1) \, \ln \, x - n \, \ln (1-x)=\ln \, 3.
\end{align}
%------------------------------------------------------------
Although these formulae (\ref{eq:curie}), (\ref{eq:ent0}), and (\ref{eq:enta}) are common to all values of $\delta$ and $n$ including the uniform case ($\delta=0$)\cite{tsh,hts}, the actual $\lambda$-dependences of these physical quantities show an infinite series of steps, 
if an infinite series of phase transition take place. 
These are plotted against $\lambda$ in  Figs. \ref{curiec} and \ref{entcr} for $\delta=\delta_{\rm c}$, where such 
 behavior is most clearly observed. These structures are reminiscent of the devil's staircase  observed in various frustrated systems and quasiperiodic systems\cite{bak}. In the case of the devil's staircase, however, physical quantities are quantized to arbitrary rational numbers and the width of the plateau decreases as the denominator increases. In contrast, in present model, the quantized values of the Curie constant and residual entropy are proportional to $1/(n+1)$ and the numerator is fixed.

\section{Summary and Discussion}

 The ground-state phases of  the alternating bond MDC with spins 1 and 1/2 are investigated. Owing to the local conservation laws,  the ground states are rigorously constructed,  once the ground states of AAFH1s are known. Each ground-state phase is described as a DC$n$ phase that consists of an uniform array of cluster-$n$'s. A cluster-$n$ is equivalent to the AAFH1 with a length of $2n+1$. 

The ground-state phase diagram in the parameter space of the frustration $\lambda$ and the distortion $\delta$ is numerically determined. 
For $\delta=\delta_{\rm c}$  $\simeq 0.2598$, 
it is also analyzed using the conformal field theory. For $\delta \simeq 1$, the cluster expansion analysis is carried out. Because both analyses predict an infinite series of phase transitions with respect to $\lambda$, we speculate that they should take place in the entire range of $\delta_{\rm c} \leq \delta \leq 1$. The behavior of the Curie constant and residual entropy is shown to exhibit  an infinite series of steps.  
 For small  $\lambda$, the ground state is the DC$\infty$ state, which corresponds to the Haldane state for $\delta < \delta_{\rm c}$ and to the dimer state for $\delta > \delta_{\rm c}$.  

Thus, we  found that the introduction of  an alternating bond distortion gives rise to a rich variety of quantum phases and phase transitions in the MDC. 
Other types of distortion that violate the local conservation laws produce even a richer variety of phases. 
These will be reported in a separate paper,%\cite{htstag} 
 where the physics of the predicted phenomena are very different from the present one.

We also stress that the possibility of the experimental realization of the MDC substantially is increased by allowing the lattice distortion. We hope the experimental synthesis of the MDC and the observation of the exotic phenomena predicted in the present paper in the near future. Variations in $\delta$ and $\lambda$ would be induced by the application of pressure. The infinite series of quantum phase transitions would manifest themselves as the pressure dependences of the low-temperature susceptibility and residual entropy.  The substitution of nonmagnetic constituent atoms would result in similar effects. Because of the massive degeneracy in the DC$n$ ground states, the magnetically ordered states would appear in the presence of an interchain interaction. Each ordered phase should have the periodicity of the corresponding DC$n$ phase in the chain direction. Such a magnetic structure is  
observable by neutron scattering experiment.

The numerical diagonalization program is based on the package TITPACK ver.2 coded by H. Nishimori.  The numerical computation in this work has been carried out using the facilities of the Supercomputer Center, Institute for Solid State Physics, University of Tokyo and the Supercomputing Division, Information Technology Center, University of Tokyo.  KH is  supported by a Grant-in-Aid for Scientific Research  on Priority Areas, "Novel States of Matter Induced by Frustration" from the Ministry of Education, Science, Sports and Culture of Japan and by a Grant-in-Aid for Scientific Research (C) from the Japan Society for the Promotion of Science. KT and HS are  supported by a Fund for Project Research from Toyota Technological Institute.


\begin{thebibliography}{33}
\bibitem{diep} {\it Frustrated Spin Systems}, ed. H. T. Diep: (World Scientific, Singapore, 2005) Chaps. 5 and 6.
\bibitem{hfm2008} {\it Proc. Int. Conf. on Highly Frustrated Magnetism (HFM2008)} J.  Phys.: Conf. Series {\bf 145} (2009).
\bibitem{mg} C. K. Majumdar and D. K. Ghosh: J. Math. Phys. {\bf 10} (1969) 1399.
\bibitem{hase} M. Hase, H. Kuroe, K. Ozawa, O. Suzuki, H. Kitazawa, G. Kido, and T. Sekine: Phys. Rev. B {\bf 70} (2004) 104426.
\bibitem{ss} B. S. Shastry and B. Sutherland: Physica B+C {\bf 108} (1981) 1069.
\bibitem{kage1} H. Kageyama, K. Yoshimura, R. Stern, N. V. Mushnikov, K. Onizuka,
M. Kato, K. Kosuge, C.P. Slichter, T. Goto, and Y. Ueda: Phys. Rev. Lett. {\bf 82} (1999) 3168.
\bibitem{kage2} H. Kageyama, M. Nishi, N. Aso, K. Onizuka, T. Yosihama, K.
Nukui, K. Kodama, K. Kakurai, and Y. Ueda: Phys. Rev. Lett. {\bf 84} (2000) 5876.
\bibitem{Takano-K-S}
K. Takano, K. Kubo, and H. Sakamoto:  
J. Phys.: Condens. Matter {\bf 8} (1996) 6405. 
\bibitem{ottk} K. Okamoto, T. Tonegawa, Y. Takahashi, and
M. Kaburagi: J. Phys.: Condens. Matter {\bf 11} (1999) 10485.
\bibitem{otk} K. Okamoto, T. Tonegawa, and M. Kaburagi: J. Phys.: Condens. Matter {\bf 15} (2003) 5979.
\bibitem{sano} K. Sano and K. Takano: J. Phys. Soc. Jpn. {\bf 69} (2000) 2710.
\bibitem{kiku} H. Kikuchi, Y. Fujii, M. Chiba, S. Mitsudo, T. Idehara, T. Tonegawa, K. Okamoto, T. Sakai, T. Kuwai, and H. Ohta: Phys. Rev. Lett. {\bf 94} (2005) 227201.
\bibitem{ohta} H. Ohta, S. Okubo, T. Kamikawa, T. Kunimoto, Y. Inagaki, H. Kikuchi, T. Saito, M. Azuma, and M. Takano: J. Phys. Soc. Jpn. {\bf 72} (2003) 2464.
\bibitem{izuoka} 
A. Izuoka, M. Fukada, R. Kumai, M. Itakura, S. Hikami, 
and T. Sugawara: 
J. Am. Chem. Soc. {\bf 116} (1994) 2609. 
\bibitem{uedia} D. Uematsu and M. Sato: J. Phys. Soc. Jpn. {\bf 76} (2007) 084712.
\bibitem{dia4spin} N. B. Ivanov, J. Richter, and J. Schulenburg : Phys. Rev. B {\bf 79} (2009) 104412.
\bibitem{str1} L. \u{C}anov\`a, J. Stre\u{c}ka, and M. Jas\u{c}\u{u}r: J. Phys.: Condens. Matter {\bf 18} (2006) 4967.
\bibitem{str2} L. \u{C}anov\`a, J. Stre\u{c}ka, and T. Lu\u{c}ivjansk\'y: Condensed Matter Phys. {\bf 12} (2009) 353.
\bibitem{fuku} H. Kobayashi, Y. Fukumoto, and A. Oguchi: J. Phys. Soc. Jpn. {\bf 78} (2009) 074004.
\bibitem{tsh}
K. Takano,  H. Suzuki, and K. Hida: 
Phys. Rev. B {\bf 80} (2009) 104410. 
\bibitem{hts}
K. Hida, K. Takano, and H. Suzuki: J. Phys. Soc. Jpn. {\bf 78} (2009) 084716.
\bibitem{plaq} N. B. Ivanov and J. Richter: Phys. Lett. A {\bf 232} (1997) 308.
\bibitem{plaq2} J. Richter, N. B. Ivanov, and J. Schulenburg: J. Phys.: Condens. Matter {\bf 10} (1998) 3635.
\bibitem{frulad1} T. Hakobyan, J. H. Hetherington, and M. Roger: Phys. Rev. B {\bf 63} (2001) 144433.
\bibitem{m-d} C. Mathoni\`ere, J.-P. Sutter, and J. V. Yakhmi: in {\it Magnetism: Molecules to Materials IV,} ed. J. S. Miller and M. Drillon: (Wiley, Weinheim, 2003) p. 1.
\bibitem{cb} Y. Hosokoshi and K. Inoue: in {\it Carbon Based Magnetism}, ed. T. L. Makarova and F. Palacio:  (Elsevier B.V., Amsterdam, 2006) p. 107.
%\bibitem{htstag}
%K. Hida, K. Takano, and H. Suzuki : in preparation. 
\bibitem{kn}
A. Kitazawa and K. Nomura: J. Phys. Soc. Jpn. {\bf 66} (1997) 3944.
\bibitem{gsh}
M. P. Gelfand, R. R. P. Singh, and D. A. Huse: J. Stat. Phys. {\bf 59} (1990) 1093.
\bibitem{bcn}
H. W. Bl\"ote, J. L. Cardy, and M. P. Nightingale: Phys. Rev. Lett. {\bf 56} (1986) 742.
\bibitem{bak} P. Bak: Rep. Prog. Phys. {\bf 45} (1982) 587.
\end{thebibliography}
\end{document}